\newif\ifdraft \draftfalse

\documentclass{CSML}

\def\dOi{12(2:7)2016}
\lmcsheading%
{\dOi}
{1--23}
{}
{}
{Jan.~13, 2014}
{Jun.~22, 2016}
{}

\ACMCCS{[{\bf Theory of computation}]: Logic---Logic and
  verification\,/\,Constructive mathematics\,/\,Type theory}

\usepackage[utf8]{inputenc}
\usepackage{color}
\usepackage{listings}
\usepackage{url}
\usepackage{breakurl}
\usepackage[breaklinks]{hyperref}
\usepackage{amssymb}
\usepackage{amsmath}
\usepackage{tikz}
\usepackage{tikz-cd}
\usetikzlibrary{arrows,matrix,backgrounds}
\usepackage{kbordermatrix}

\theoremstyle{plain}
\theoremstyle{remark}

\newcommand{\Coq}{{\sc Coq}}
\newcommand{\ssr}{{\sc SSReflect}}
\newcommand{\CoqEAL}{{\sc CoqEAL}}
\newcommand{\Haskell}{{\sc Haskell}}
\newcommand{\Mizar}{{\sc Mizar}}
\newcommand{\Isabelle}{{\sc Isabelle}}
\newcommand{\HOLLight}{{\sc HOL Light}}
\newcommand{\ACL}{{\sc ACL2}}
\newcommand{\C}[1]{\mbox{\lstinline`#1`}}
\newcommand\N[1]{\langle\mbox{\itshape\rmfamily\small #1}\rangle}
\newcommand\Z{\mathbb{Z}}
\renewcommand\N{\mathbb{N}}

\newcommand{\eg}{{\em e.g.}}
\newcommand{\ie}{{\em i.e.}}
\newcommand{\cf}{{\em cf.}}
\newcommand{\lipsis}{$\ldots$}

\let\L=\lstinline

\definecolor{dkblue}{rgb}{0,0.1,0.5}
\definecolor{lightblue}{rgb}{0,0.5,0.5}
\definecolor{dkgreen}{rgb}{0,0.4,0}
\definecolor{dk2green}{rgb}{0.4,0,0}
\definecolor{dkviolet}{rgb}{0.6,0,0.8}
\definecolor{shadethmcolor}{rgb}{0.9, 0.9,1}

\font\eightrm=cmr8
\def\eqd{\hbox{ \eightrm \%}\kern -3pt =}

\let\amsamp=&

\begingroup
\catcode`\&=13
\gdef\pampmatrix{%
  \begingroup
  \let&=\amsamp
  \begin{pmatrix}%
}
\gdef\endpampmatrix{\end{pmatrix}\endgroup}
\endgroup

\definecolor{dkblue}{rgb}{0,0.1,0.5}
\definecolor{dkgreen}{rgb}{0,0.6,0}
\definecolor{dkred}{rgb}{0.6,0,0}
\definecolor{dkpurple}{rgb}{0.7,0,0.4}

\renewcommand{\sec}[1]{section~\ref{#1}}
\newcommand{\fig}[1]{figure~\ref{#1}}

\pgfdeclarelayer{myback}
\pgfsetlayers{myback,background,main}

\tikzset{mycolor/.style = {line width=1bp,color=#1}}%
\tikzset{myfillcolor/.style = {draw,fill=#1}}%

\newcommand{\fhighlight}[3]{%
\draw[myfillcolor=#1] (#2.north west)rectangle (#3.south east);
}

\DeclareMathOperator{\coker}{coker}
\DeclareMathOperator{\gdco}{gdco}
\DeclareMathOperator{\im}{\mathcal{I}m}

\begin{document}

\title[Elementary Divisor Rings in Coq]{Formalized Linear Algebra over
  Elementary Divisor Rings in Coq}

\author[G.~Cano]{Guillaume Cano\rsuper a}
\address{{\lsuper a}University of Perpignan}
\email{guillaume.cano@univ-perp.fr}

\author[C.~Cohen]{Cyril Cohen\rsuper b}
\address{{\lsuper b}Inria Sophia Antipolis -- Méditerranée}
\email{cyril.cohen@inria.fr}

\author[M.~Dénès]{Maxime Dénès\rsuper c}
\address{{\lsuper c}University of Pennsylvania}
\email{mail@maximedenes.fr}

\author[A.~Mörtberg]{Anders Mörtberg\rsuper d}
\address{{\lsuper d}University of Gothenburg}
\email{anders.mortberg@cse.gu.se}

\author[V.~Siles]{Vincent Siles\rsuper e}
\address{{\lsuper d}University of Gothenburg}
\email{vincent.siles@ens-lyon.org}

%% mandatory lists of keywords and classifications:
% \subjclass{MANDATORY list of acm classifications}
% Found at: http://delivery.acm.org/10.1145/2380000/2371137/ACMCCSTaxonomy.html?ip=129.16.22.67&id=2371137&acc=OPEN&key=74F7687761D7AE37.3C5D6C4574200C81.4D4702B0C3E38B35.6D218144511F3437&CFID=424525785&CFTOKEN=76268892&__acm__=1395310295_6d987107b53bbb0082b5b8fe4998f060#_Toc320886570

% \titlecomment{OPTIONAL comment concerning the title, \eg, if a variant
% or an extended abstract of the paper has appeared elsewehere}

\begin{abstract}
  This paper presents a \Coq{} formalization of linear algebra over
  elementary divisor rings, that is, rings where every matrix is
  equivalent to a matrix in Smith normal form. The main results are
  the formalization that these rings support essential operations of
  linear algebra, the classification theorem of finitely presented
  modules over such rings and the uniqueness of the Smith normal form
  up to multiplication by units. We present formally verified
  algorithms computing this normal form on a variety of coefficient
  structures including Euclidean domains and constructive principal
  ideal domains. We also study different ways to extend Bézout domains
  in order to be able to compute the Smith normal form of
  matrices. The extensions we consider are: adequacy (\ie{} the
  existence of a~$\gdco$ operation), Krull dimension~$\leq 1$ and
  well-founded strict divisibility.
\end{abstract}

\maketitle

% \tableofcontents

\section{Introduction}\label{sec:introduction}

The goal of this paper is to develop linear algebra for
\emph{elementary divisor rings}, that is, rings where there is an
algorithm for computing the Smith normal form of matrices. More
specifically, we focus on the axiomatics and basic algorithms of such
rings. This work fits within a bigger project of formalizing theories
and algorithms for constructive module theory. This lays ground to
program more efficient algorithms (like in {\sc Axiom}, {\sc Maple},
{\sc Magma}, \lipsis), and prove them correct with regard to the
notions introduced here, using the \CoqEAL{} methodology
\cite{coqeal,refinements_for_free}, developed by some of the authors
(\cf{} \sec{sec:related}).

In this paper, we do not contribute with new theorems or new efficient
algorithms in the field, except for some minor factoring in order to
simplify the formal proof. However, we provide a formal framework to
develop more theory and more complex algorithms in \Coq{}, by
regrouping and representing folklore concepts in the proof assistant,
that we found scattered in the literature.  We make a synthesis of
different axiomatics and show how they are linked constructively. In
order to do that, we first introduce the classical notions at use and
their constructive variants (\sec{sec:class-axiom-constr}).

The main contributions of this paper are the
formalization\footnote{The formal development is a subset of the
  repository: \url{https://github.com/CoqEAL/CoqEAL}, and the
  companion material for this paper has been regrouped here:
  \url{http://www.cyrilcohen.fr/work/edr/}}, using the \Coq{} proof
assistant with the \ssr{} extension, of:

\begin{itemize}

\item rings with explicit divisibility, GCD domains, Bézout domains,
  constructive principal ideal domains and Euclidean
  domains~(\sec{sec:dvdrings}), which corresponds to the file
  {\tt theory/dvdring.v};

\item an algorithm computing the Smith normal form of matrices with
  coefficients in Euclidean domains (files {\tt theory/dvdring.v} and
  {\tt refinements/smith.v}) and the generalization to constructive
  principal ideal domains~(\sec{sec:algo}), which corresponds to the file
  {\tt refinements/smithpid.v};

\item linear algebra over elementary divisor rings together with a
  proof that the Smith normal form is unique up to multiplication by
  units for rings with a~$\gcd$ operation~(\sec{sec:edr}) (file {\tt
    theory/edr.v}) and the classification theorem for finitely
  presented modules over elementary divisor rings (file {\tt
    theory/fpmod.v}); and

\item proofs that Bézout domains extended with one of the three
  extensions above are elementary divisor rings and how these notions
  are related~(\sec{sec:kaplansky}) which corresponds to the file {\tt
    theory/kaplansky.v}.
\end{itemize}
The paper ends with an overview of related work~(\sec{sec:related}), followed by
conclusions and future work~(\sec{sec:conclusion}).

\section{Classical axiomatics and constructive variants}
\label{sec:class-axiom-constr}

The algorithms we present for computing the Smith normal form can be
seen as generalizations of Gaussian elimination that can, in
particular, be defined for $\Z$. The main source of
inspiration for this work is the formalization of finite dimensional
vector spaces by Georges Gonthier~\cite{gonthier_point-free_2011} in
which spaces are represented using matrices and all subspace
constructions can be elegantly defined from Gaussian elimination. This
enables a concrete and point-free presentation of linear algebra which
is suitable for formalization as it takes advantage of the \emph{small
  scale reflection} methodology of the \ssr{} extension and the
Mathematical Components library~\cite{SSReflect} for the \Coq{} proof
assistant~\cite{Coq}. When generalizing this to elementary divisor
rings there are two essential problems that need to be resolved before
the theory may be formalized:

\begin{enumerate}
\item What is a suitable generalization of finite dimensional vector
  spaces when considering more general classes of rings than fields as
  coefficients?

\item What rings are elementary divisor rings?
\end{enumerate}
A possible answer to the first problem is finitely generated
$R$-modules, \ie~finite dimensional vector spaces with coefficients in
a general ring instead of a field.
% However as we want to represent all structures concretely using
% matrices in order to utilize that we have an algorithm for computing
% the Smith normal form \md{Is this the real reason? Or is it that
% some algorithmic problems are inherently not solvable for general
% modules?}  this is not a suitable answer. For vector spaces it is
% possible to restrict the attention to those that are finite
% dimensional and utilize that any such vector space has a basis, this
% suggest that we should restrict our attention to \emph{finitely
% generated} modules.
However these are not as well behaved as finite dimensional vector
spaces as there might be relations among the generators. In other
words, not all finitely generated modules are \emph{free}. To overcome
this, we restrict our attention further and consider \emph{finitely
  presented} modules, which are modules specified by a finite number
of generators and a finite number of relations between these. This
class of modules may be represented concretely using matrices, which
in turn means that we can apply the same approach as
in~\cite{gonthier_point-free_2011} and implement all operations by
manipulating the presentation matrices.

A standard answer to the second problem is \emph{principal ideal
  domains} like $\Z$ and the ring of univariate polynomials over a
field (denoted by~$k[x]$). The classical definition of principal ideal
domains is integral domains where \emph{all} ideals are principal
(\ie{} generated by one element). In particular it means that
principal ideal domains are \emph{Noetherian} as all ideals are
finitely generated. Classically this is equivalent to the ascending
chain condition for ideals, however in order to prove this equivalence
classical reasoning is used in essential ways. In fact, if these
definitions are read constructively they are so strong that no ring
except the trivial ring satisfies them~\cite{Perdry2004511}. Principal
ideal domains are hence problematic from a constructive point of view
as they are Noetherian.
% , however we will show that it is possible to give a constructive
% version of the ascending chain condition in type theory that allows
% us to define a constructive approximation of them.

A possible solution is to restrict the attention to Euclidean domains
(which include both~$\Z$ and~$k[x]$) and show how to compute the Smith
normal form of matrices over these rings. This approach is appealing
as it allows for a simple definition of the Smith normal form
algorithm that resembles the one of Gaussian elimination. While
Euclidean domains are important, we would like to be more general. In
order to achieve this we consider an alternative approach that is
customary in constructive algebra: to generalize all statements and
not assume Noetheriannity at all~\cite{LomQui}. If we do this for
principal ideal domains we get \emph{Bézout domains}, which are rings
where every \emph{finitely generated} ideal is principal. However, it
is an open problem whether all Bézout domains are elementary divisor
rings or not~\cite{Lorenzini}. Hence we study different assumptions
that we can add to Bézout domains in order to prove that they are
elementary divisor rings. The properties we define and study
independently are:

\begin{enumerate}
  \item Adequacy (\ie{} the existence of a $gdco$ operation);
  \item Krull dimension $\leq 1$;
  \item Strict divisibility is well-founded.
\end{enumerate}
The last one can be seen as a constructive approximation to the
ascending chain condition for principal ideals, so this kind of Bézout
domains will be referred to as \emph{constructive principal ideal
  domains}.

\section{Rings with explicit divisibility}\label{sec:dvdrings}
In this section we recall definitions and basic properties of rings
with explicit divisibility, GCD domains, Bézout domains, constructive
principal ideal domains and Euclidean domains.

\subsection{Rings with explicit divisibility}

Throughout the paper all rings are discrete integral domains, \ie{}
commutative rings with a unit, decidable equality and no zero
divisors. This section is loosely based on the presentation of
divisibility in discrete domains of Mines, Richman and Ruitenberg
in~\cite{Mines}. The central notion we consider is:

\begin{defi}
  A ring $R$ has \textbf{explicit divisibility} if it has a
  divisibility test that produces witnesses.
\end{defi}

That is, given $a$ and $b$ we can test if $a \mid b$ and if this is the case get
$x$ such that $b = xa$. Two elements $a,b \in R$ are \emph{associates} if $a
\mid b$ and $b \mid a$, which is equivalent to $b = ua$ for some unit $u$
because we have cancellation. Note that this gives rise to an equivalence
relation. This notion will play an important role later as we will show that the
Smith normal form of a matrix is unique up to multiplication by units, that is,
up to associated elements.

A GCD domain is an example of a ring with explicit divisibility:

\begin{defi}
  A \textbf{GCD domain} $R$ is a ring with explicit divisibility in
  which every pair of elements has a greatest common divisor, that is,
  for $a, b \in R$ there is $\gcd(a,b)$ such that $\gcd(a,b) \mid a$,
  $\gcd(a,b) \mid b$ and $\forall g,\, (g \mid a) \land (g \mid b)
  \rightarrow g \mid \gcd(a,b) $.
\end{defi}

Note first that we make no restriction on $a$ and $b$, so they can
both be zero. In this case the greatest common divisor is zero. This
makes sense as zero is the maximum element for the divisibility
relation. Note also that as $R$ is assumed to be a ring with explicit
divisibility we get that $\gcd(a,b) \mid a$ means that there is $a'$
such that $a = a' \gcd(a,b)$. By Euclid's algorithm we know that both
$\Z$ and $k[x]$ are GCD domains.

With the above definition the greatest common divisor of two elements
is not necessarily unique, \eg{} the greatest common divisor of $2$
and $3$ in $\Z$ is either $1$ or $-1$. But if we consider equality up
to multiplication by units (\ie{} up to associatedness) the greatest
common divisor is unique, so in the rest of the paper equality will
denote equality up to associatedness when talking about the $\gcd$.

Most of the rings we will study in this paper are Bézout domains:

\begin{defi}
  A \textbf{Bézout domain} is a GCD domain $R$ such that for any two
  elements $a,b \in R$ there is $x,y \in R$ such that $ax + by =
  \gcd(a,b)$.
\end{defi}

Let $a$ and $b$ be two elements in a ring $R$. If $R$ is a GCD domain
we can compute $g = \gcd(a,b)$ together with witnesses to the ideal
inclusion $(a,b) \subseteq (g)$. Further, if $R$ is a Bézout domain we
can compute witnesses for the inclusion $(g) \subseteq (a,b)$ as
well. This can be generalized to multiple elements
$a_1,\dots,a_n \in R$ to obtain witnesses for the inclusions
$(a_1,\dots,a_n) \subseteq (g)$ and $(g) \subseteq (a_1,\dots,a_n)$
where $g$ is the greatest common divisor of the~$a_i$. Bézout domains
can hence be characterized as rings in which every finitely generated
ideal is principal, which means that they are non-Noetherian
generalizations of principal ideal domains.

Note that, on the one hand there exists $a'$ and $b'$ such that
$a = a' g$ and $b = b' g$, and on the other hand we have $x$ and $y$
such that $ax+by = g$. Therefore, by dividing with $g$, we obtain a
Bézout relation between $a'$ and $b'$, namely $a'x+a'y = 1$.

This definition can be extended to give a constructive version of
principal ideal domains. We say that $a$ divides $b$ \emph{strictly}
if $a \mid b$ but $b \nmid a$, then we can define:

\begin{defi}
  A \textbf{constructive principal ideal domain} is a Bézout domain in
  which the strict divisibility relation is well-founded.
\end{defi}

By well-founded we mean that any descending chain of strict divisions
is finite. This can be seen as a constructive approximation to the
ascending chain condition for principal ideals and hence to
Noetheriannity. Both $\Z$ and $k[x]$ can be proved to be Bézout
domains and satisfy the condition of constructive principal ideal
domains. In fact, this can be done for any ring on which the extended
Euclidean algorithm can be implemented. These rings are called
Euclidean domains:

\begin{defi}
  A \textbf{Euclidean domain} is a ring $R$ with a Euclidean norm
  $\mathcal{N} : R \rightarrow \N$ such that for any $a \in R$ and
  nonzero $b \in R$ we have $\mathcal{N}(a) \leqslant \mathcal{N}(ab)
  $.  Further, for any $a \in R$ and nonzero $b \in R$ we can find
  $q,r \in R$ such that $a = bq + r$ and either $r = 0$ or
  $\mathcal{N}(r) < \mathcal{N}(b)$.
\end{defi}

In the case of $\Z$ and $k[x]$ we can take respectively the absolute value
function and the degree function as Euclidean norm. Then the standard
division algorithms for these rings can be used to compute $q$ and
$r$.

\subsection{Formalization of algebraic structures}\label{sec:formal}
The algebraic structures have been formalized in the same manner as in
the \ssr~library~\cite{SSReflect-hierarchy} using packed classes
(implemented by mixins and canonical structures). We will now discuss
the formalization of these new structures starting with the definition
of rings with explicit divisibility:

\begin{lstlisting}
Inductive div_spec (R : ringType) (a b :R) : option R -> Type :=
  | DivDvd x of a = x * b : div_spec a b (Some x)
  | DivNDvd of (forall x, a != x * b) : div_spec a b None.

Record mixin_of R := Mixin {
  div : R -> R -> option R;
  _ : forall a b, div_spec a b (div a b)
}.
\end{lstlisting}
This structure is denoted by \L{DvdRing} and for a ring to be an
instance it needs to have a function \L{div} that returns an
\L{option} type, such that if \L{div a b = None} then $a \nmid b$, and
if \L{div a b = Some x} then $x$ is the witness that $a \mid b$. The
notation used for \L{div a b} in the formalization is
\L{a %/? b}. There is also a \L{%|} notation for the
  \L{div} function that returns a boolean, this relies on a coercion
  from \L{option} to \L{bool} defined in the \ssr~libraries (mapping
  \L{None} to \L{false} and \L{Some x} to \L{true} for any \L{x}).
  Using this we have implemented the notion of associatedness, denoted
  by \L{%=}, and the basic theory of divisibility.

Next we have the \L{GCDDomain} structure which is implemented as:

\begin{lstlisting}
Record mixin_of R := Mixin {
  gcd : R -> R -> R;
  _ : forall d a b, (d %| gcd a b) = (d %| a) && (d %| b)
}.
\end{lstlisting}
For a ring to be a \L{GCDDomain} it needs to have a $\gcd$ function
satisfying the property above. This property is sufficient as it
implicitly gives that $\gcd(a,b) \mid a$ and $\gcd(a,b) \mid b$ since
divisibility is reflexive.

The \L{BezoutDomain} structure looks like:

\begin{lstlisting}
Inductive bezout_spec (R : gcdDomainType) (a b : R) : R * R -> Type :=
  BezoutSpec x y of gcdr a b %= x * a + y * b : bezout_spec a b (x, y).

Record mixin_of R := Mixin {
  bezout : R -> R -> R * R;
   _ : forall a b, bezout_spec a b (bezout a b)
}.
\end{lstlisting}
Recall that a constructive principal ideal domain is a Bézout domain
where strict divisibility is well-founded. This is denoted by \L{PID}
and is implemented by:

\begin{lstlisting}
Definition sdvdr (R : dvdRingType) (x y : R) := (x %| y) && ~~(y %| x).
\end{lstlisting}

\begin{lstlisting}
Record mixin_of R := Mixin {
  _ : well_founded (@sdvdr R)
}.
\end{lstlisting}
The notation \L+x %<| y+ will be used to denote \L+sdvdr x y+
and \L+~~+ denotes the boolean negation.
We will see more precisely in \sec{sec:smith-pid} how \L{well_founded}
is defined formally in \Coq{}'s standard library when we use it to
prove the termination of our Smith normal form algorithm.

We also have the \L{EuclideanDomain} structure that represents Euclidean
domains:

\begin{lstlisting}
Inductive edivr_spec (R : ringType)
  (g : R -> nat) (a b : R) : R * R -> Type :=
  EdivrSpec q r of a = q * b + r & (b != 0) ==> (g r < g b)
  : edivr_spec g a b (q, r).

Record mixin_of R := Mixin {
  enorm : R -> nat;
  ediv : R -> R -> R * R;
  _ : forall a b, a != 0 -> enorm b <= enorm (a * b);
  _ : forall a b, edivr_spec enorm a b (ediv a b)
}.
\end{lstlisting}
This structure contains the Euclidean norm and the Euclidean division function
together with their proofs of correctness. We have implemented the extended
version of Euclid's algorithm for Euclidean domains and proved that it satisfies
\L{bezout_spec}. Hence we get that Euclidean domains are Bézout domains. We have
also proved that any \L{EuclideanDomain} is a \L{PID} which means that strict
divisibility is well-founded in both $\Z$ and $k[x]$.

The relationship between the algebraic structures presented in this
section can be depicted by:
\[
\text{\L{EuclideanDomain}} \subset
\text{\L{PID}}             \subset
\text{\L{BezoutDomain}}    \subset
\text{\L{GCDDomain}}       \subset
\text{\L{DvdRing}}         \subset
\text{\L{IntegralDomain}}
\]
where \L{IntegralDomain} is already present in the \ssr~hierarchy. In
the next section we consider an algorithm for computing the Smith
normal form of matrices over the first two algebraic structures in the
chain of inclusions. This means that these two structures are
elementary divisor rings. In~\sec{sec:kaplansky} we will generalize to
Bézout domains of Krull dimension $\leq 1$ and adequate domains that
fit in between \L{PID} and \L{BezoutDomain} in the chain of
inclusions.

%% Local Variables:
%% ispell-local-dictionary: "english"
%% mode: latex flyspell
%% TeX-master: "main"
%% End:

\section{A verified algorithm for the Smith Normal Form}\label{sec:algo}

In \cite{Kaplansky} Kaplansky introduced the notion of
\textbf{elementary divisor rings} as rings where every matrix is
equivalent to a matrix in Smith normal form, that is, given a
$m\times n$ matrix $M$ there exist invertible matrices $P$ and $Q$ of
size $m \times m$ and $n \times n$ respectively, such that $PMQ = D$
where $D$ is a diagonal matrix of the form:

\[
\begin{bmatrix}
  d_1    &        &  0      & \cdots & \cdots & 0      \\
         & \ddots &         &        &        & \vdots \\
  0      &        & d_k     & 0      & \cdots & 0      \\
  \vdots &        & 0       & 0      &        & \vdots \\
  \vdots &        & \vdots  &        & \ddots & \vdots \\
  0      & \cdots & 0       & \cdots & \cdots & 0      \\
\end{bmatrix}
\]

\noindent with the additional property that $d_i~|~d_{i+1}$ for all
$i$. % the < k was not necessary as anything divide 0

Let us first explain how we formalized the notion of Smith normal form
in \Coq{}, with the following representation of matrices taken from
the \ssr{} library:
\begin{lstlisting}
Inductive matrix R m n := Matrix of {ffun 'I_m * 'I_n -> R}.
\end{lstlisting}
Here \L{'I_m} is the type of ordinals (\ie~natural numbers bounded by
\L{m}) which has exactly \L{m} inhabitants and can be coerced to
\L{nat}. Matrices are then implemented as finite functions over finite
sets of indices, with dependent types being used to ensure
well-formedness. We use the notation \L{'M[R]_(m,n)} for the type
\L{matrix R m n}, the notation \L{'rV[R]_m} for the type of row
vectors of length \L{m} and the notation \L{'cV[R]_m} for column
vectors of height \L{m}. The ring \L{R} is often omitted from these
notations when it can be inferred from the context.

In order to express that a matrix is in Smith normal form, we define
\L{diag_mx_seq}, which rebuilds a diagonal matrix from a list~(note
that the type of lists is called \L{seq} in the \ssr~library) of
diagonal coefficients:
\begin{lstlisting}[language=SSR]
Definition diag_mx_seq m n (s : seq R) :=
  \matrix_(i < m, j < n) s`_i *+ (i == j :> nat).
\end{lstlisting}
The notation \L{x *+ n}, where \L{x} belongs to a ring and \L{n} is a
natural number, stands for the sum \L{x $+ \ldots +$ x} iterated \L{n}
times. In the expression of the general coefficients of the matrix
above, \L{i} and \L{j} are ordinals of type \L{'I_m} and \L{'I_n}
respectively.  The notation \L{i == j :> nat} tells \Coq{} to compare
them as natural numbers and returns a boolean. A coercion then sends
this boolean to a natural number (\L{true} is interpreted by \L{1} and
\L{false} by \L{0}). Thus \L{s`_i *+ (i == j :> nat)} denotes the
element of index \L{i} in \L{s} if \L{i} and \L{j} have the same
value, \L{0} otherwise.

Now if \L{M} is a matrix, an algorithm for computing the Smith normal
form should return a list \L{s} and two matrices \L{P} and \L{Q} such
that:

\begin{itemize}
\item The sequence \L{s} is sorted for the divisibility relation.
\item The matrix \L{diag_mx_seq m n s} is equivalent to \L{M}, with
  transition matrices~\L{P} and~\L{Q}.
\end{itemize}

\noindent Which translates formally to an inductive predicate:
\begin{lstlisting}[language=SSR]
Inductive smith_spec R m n M : 'M[R]_m * seq R * 'M[R]_n -> Type :=
  SmithSpec P d Q of P *m M *m Q = diag_mx_seq m n d
                     & sorted %| d
                     & P \in unitmx
                     & Q \in unitmx : smith_spec M (P,d,Q).
\end{lstlisting}
We have packaged this in the same manner as above in order to
represent elementary divisor rings: 

\begin{lstlisting}[language=SSR]
Record mixin_of R := Mixin {
  smith : forall m n, 'M[R]_(m,n) -> 'M[R]_m * seq R * 'M[R]_n;
  _ : forall m n (M : 'M[R]_(m,n)), smith_spec M (smith M)
}.
\end{lstlisting}

\noindent In the rest of this section we will see direct proofs that
Euclidean domains and constructive principal ideal domains provide
instances of this structure.

\subsection{Smith normal form over Euclidean domains}\label{sec:euclidean}
We mentioned in the introduction that constructive finite dimensional linear
algebra over a field can be reduced to matrix encodings. Information like the
rank and determinant is then reconstructed from the encoding using Gaussian
elimination, which involves three kinds of operations on the matrix:
\begin{enumerate}
  \item Swapping two rows (resp. columns)
  \item Multiplying one row (resp. column) by a nonzero constant
  \item Adding to a row (resp. column) the product of another one by a constant
\end{enumerate}

These three operations are interesting because they are compatible with matrix
equivalence. In particular, they can be expressed as left (resp. right)
multiplication by invertible matrices.

The same algorithm fails to apply in general to a matrix over a ring, since it
may require a division by the pivot, which could be not exact. The content of
this section can thus be seen as a generalization of Gaussian elimination to
Euclidean domains.

To make this extension possible, a new kind of elementary operations
needs to be introduced. Let $a$ and $b$ be elements of a Euclidean
domain $R$.  Bézout's identity gives $u$ and $v$ such that $ua + vb =
\gamma$ where $\gamma = \gcd(a,b)$. Let us note $a' =
\frac{a}{\gamma}$ and $b'= \frac{b}{\gamma}$, these divisions being
exact by definition of the $\gcd$. We get the identity: $ua' + vb' =
1$. Consider the following square matrix of size $n$:

\[ E_{\mathrm{Bezout}}(a,b,n,k) = \kbordermatrix{
  &  &   &        &   & (\textrm{col. }k) \\
&u   &   &        &   & v \\
&    & 1 & \\
&    &   & \ddots & \\
&    &   &        & 1 & \\
(\textrm{row }k) &-b' &   &        &   & a' \\
&    &   &        &   &   & 1 \\
&    &   &        &   &   &   & \ddots \\
&  &   &        &   &   &   &        & 1}
\]
The coefficients not explicitly shown in this matrix
%$E_{\mathrm{Bezout}}$
 are
assumed to be zeros. Note that
$\det(E_{\mathrm{Bezout}}(a,b,n,k)) = ua' + vb' = 1$, so in particular
the matrix above is invertible.

We formalize these matrices as follows:

\begin{lstlisting}[language=SSR]
Definition combine_mx (a b c d : R) (m : nat) (k : 'I_m) :=
  let k' := lift 0 k in
  let d := \row_j (a *+ (j == 0) + d *+ (j == k') +
                    ((j != 0) && (j != k'))%:R) in
  diag_mx d + c *: delta_mx k' 0 + b *: delta_mx 0 k'.

Definition Bezout_mx (a b : R) (m : nat) (k : 'I_m) :=
  let:(_,u,v,a1,b1) := egcdr a b in combine_mx u v (-b1) a1 k.
\end{lstlisting}
For an ordinal \L{i} of type \L{'I_m}, \L{lift 0 i} represents the
ordinal \L{1 + i} of type \L{'I_(1 + m)}. The notation %
\L{\row_(j < m) r j} corresponds to the row matrix %
\L{[r 0, ..., r (m-1)]}, if the dimension can be automatically
inferred then we can just write \L{\row_j r j}.
If \L{b} is a boolean, the term \L{b%:R} reduces to \L{1} if \L{b} is true,
\L{0} otherwise. The matrix \L{diag_mx d} correspond to the diagonal matrix
where diagonal coefficients are the coefficients of the row matrix \L{d}, and
\L{delta_mx i j} is the matrix which has only zeros except at position $(i,j)$,
where the coefficient is \L{1}. Finally, \L{a *: A} is the matrix \L{A} multiplied
by the scalar \L{a}. Note that the Bézout identity between \L{a} and \L{b} is
given by the function \L{egcdr}, which is exported by the underlying Euclidean
ring.

Like other elementary operations, multiplication by
$E_{\mathrm{Bezout}}(a,b,n,k)$ on the left corresponds to an operation
on the rows:
\[
  E_{\mathrm{Bezout}}(a,b,n,k) \times \begin{bmatrix}
    L_1\\
    L_2\\
    \vdots\\
    L_{k-1}\\
    L_k\\
    L_{k+1}\\
    \vdots\\
    L_n
  \end{bmatrix} = \begin{bmatrix}
    u L_1 + v L_k\\
    L_2\\
    \vdots\\
    L_{k-1}\\
    -b' L_1 + a' L_k\\
    L_{k+1}\\
    \vdots\\
    L_n
  \end{bmatrix}
\]
These row operations are described formally by:
\begin{lstlisting}[language=SSR]
Definition combine_step (a b c d : R) (m n : nat)
                           (M : 'M_(1 + m,1 + n)) (k : 'I_m) :=
  let k' := lift 0 k in
  let r0 := a *: row 0 M + b *: row k' M in
  let rk := c *: row 0 M + d *: row k' M in
  \matrix_i (r0 *+ (i == 0) + rk *+ (i == k') +
              row i M *+ ((i != 0) && (i != k'))).

Definition Bezout_step (a b : R) (m n : nat)
                          (M : 'M_(1 + m, 1 + n)) (k : 'I_m) :=
  let:(_,u,v,a1,b1) := egcdr a b in combine_step u v (-b1) a1 M k.
\end{lstlisting}
Here \L{row i M} represents the $i$:th row of \L{M}. A lemma connects
these row operations to the corresponding elementary matrices:

\begin{lstlisting}[language=SSR]
Lemma Bezout_stepE a b (m n : nat) (M : 'M_(1 + m,1 + n)) k :
  Bezout_step a b M k = Bezout_mx a b k *m M.
\end{lstlisting}
Let now $M=(a_{i,j})$ be a matrix with coefficients in $R$. We will now show how
to reduce~$M$ to its Smith normal form using elementary operations.
As for Gaussian elimination, we start by finding a nonzero pivot $g$ in $M$,
which is moved to the upper-left corner~(if~$M=0$,~$M$ is in Smith normal
form). We search the first column for an element which is not divisible by
$g$. Let us assume that $g \nmid a_{k,1}$, we then multiply the matrix on the
left by $E_{\mathrm{Bezout}}(g,a_{k,1},n,k)$~:
\[ E_{\mathrm{Bezout}}(g,a_{k,1},n,k) \times
  \begin{bmatrix}
    g & L_1 \\
    a_{2,1} & L_2 \\
    \vdots & \vdots\\
    a_{k,1} & L_k \\
    \vdots & \vdots \\
    a_{n,1} & L_n\\
  \end{bmatrix} =
\begin{bmatrix}
    \gamma & u L_1 + v L_k\\
    a_{2,1} & L_2\\
    \vdots\\
    -g' g + a' a_{k,1} & -g' L_1 + a' L_k\\
    \vdots & \vdots\\
    a_{n,1} & L_n
\end{bmatrix} \]
with the Bézout identity $u g + v a_{k,1} = \gamma = \gcd(g,a_{k,1})$ and posing
as previously $g' = \frac{g}{\gamma}$, we have $a' = \frac{a_{k,1}}{\gamma}$.

By definition of $\gamma$, we have: $\gamma \mid -g' g + a'
a_{k,1}$.
Moreover, all the coefficients in the first column of $M$ which were divisible
by $g$ are also by $\gamma$. We can therefore repeat this process until we get a
matrix whose upper-left coefficient (which we still name $g$) divides all the
coefficients in the first column. Linear combinations on rows can thence lead to
a matrix $B$ of the following shape:

\[B =
\begin{tikzpicture}[baseline=-\the\dimexpr\fontdimen22\textfont2\relax ]
  \matrix (m)[matrix of math nodes,left delimiter=\lbrack{},right
  delimiter=\rbrack{}]
{
  g & b_{1,2} & \cdots & b_{1,n} \\
g & \vdots &  & \vdots\\
\vdots & \vdots &  & \vdots\\
g & b_{m,2} & \cdots & b_{m,n} \\
};

\begin{pgfonlayer}{myback}
  \fhighlight{blue!40}{m-1-2}{m-4-4}
\end{pgfonlayer}
\end{tikzpicture}
\]\smallskip

\noindent We then search the indicated submatrix of $B$ for an element that is
not divisible by $g$. If such a coefficient $b_{i,j}$ is found, it is
moved to the top by permuting rows $1$ and $i$. Thus $g$ is still the
upper-left coefficient\footnote{The authors were inspired by the use
  of a similar trick in an algorithm formalized by Georges Gonthier.}
and multiplications on the right by $E_{\mathrm{Bezout}}$ matrices
allow, like previously, to obtain a matrix whose upper-left
coefficient divides all the others.

This first step is implemented by the function \L{improve_pivot_rec}:
\begin{lstlisting}[language=SSR,numbers=left,firstnumber=1]
Fixpoint improve_pivot_rec k {m n} :
  'M[R]_(1 + m) -> 'M[R]_(1 + m, 1 + n) -> 'M[R]_(1 + n) ->
  'M[R]_(1 + m) * 'M[R]_(1 + m, 1 + n) * 'M[R]_(1 + n) :=
  match k with
  | 0 => fun P M Q => (P,M,Q)
  | p.+1 => fun P M Q =>
      let a := M 0 0 in
      if find1 M a is Some i then
        let Mi0 := M (lift 0 i) 0 in
        let P := Bezout_step a Mi0 P i in
        let M := Bezout_step a Mi0 M i in
        improve_pivot_rec p P M Q
      else
      let u  := dlsubmx M in let vM := ursubmx M in let vP := usubmx P in
      let u' := map_mx (fun x => 1 - odflt 0 (x %/? a)) u in
      let P  := col_mx (usubmx P) (u' *m vP + dsubmx P) in
      let M  := block_mx a%:M vM
                         (const_mx a) (u' *m vM + drsubmx M) in
      if find2 M a is Some (i,j) then
        let M := xrow 0 i M in let P := xrow 0 i P in
        let a := M 0 0 in
        let M0ij := M 0 (lift 0 j) in
        let Q := (Bezout_step a M0ij Q^T j)^T in
        let M := (Bezout_step a M0ij M^T j)^T in
        improve_pivot_rec p P M Q
      else (P, M, Q)
  end.
\end{lstlisting}
If $A$, $B$, $C$ and $D$ are four matrices (with matching
dimensions) then \L{block_mx A B C D} is the matrix:

\[M = \begin{bmatrix} A & B \\ C & D \end{bmatrix}\]
where \L{A = ulsubmx M}, \L{B = ursubmx M}, \L{C = dlsubmx M} and %
\L{D = drsubmx M}. Similarly \L{C = col_mx A B} is a column matrix
with \L{A = usubmx C} and \L{B = dsubmx C} (the functions for
constructing and destructing row matrices have similar names). The
matrix \L{const_mx a} is the matrix where each coefficient is equal to
\L{a} and %
\L{xrow i j M} is the matrix \L{M} with the rows \L{i} and \L{j}
exchanged.

The function \L{improve_pivot_rec} takes as arguments a natural number \L{k}
which represents the number of remaining steps, the original matrix and two
current transition matrices. If the number of remaining steps is zero, the
matrices are returned unchanged (line~5). If not, the first column is searched
for an element that is not divisible by the pivot (function \L{find1},
line~8). If such an element is found on a row of index \L{i}, a Bézout step is
performed between the first row and the one of index \L{i}, and the function is
called recursively (lines~9 to~12). If, on the contrary, the pivot divides all
the elements in the first column, some linear combinations (lines~14 to~18)
bring us back to a matrix of the shape of the matrix $B$ seen above. Finally,
the remaining lines search the whole matrix for an element that is not divisible
by the pivot (function \L{find2}), perform a Bézout step on the columns if
appropriate, and call the function recursively.

We have made several choices when implementing this function. First, the
argument \L{k} bounding the number of steps makes it easy to have a
structural recursion (this natural number decreases by $1$ at each step). In
this usual technique, \L{k} is often called the fuel of the recursion. The flip
side is that in order to call the function, an a priori bound on the number of
steps has to be provided. It is at this point that the hypothesis we made that
$R$ is a Euclidean domain comes in handy: we can take as a bound the Euclidean
norm of the upper-left coefficient of the original matrix.

We also chose to abstract over initial transition matrices, which are updated as
the process goes on. From a computational standpoint, this approach has two
benefits. First, it avoids the need for products by transition matrices,
asymptotically more costly than to perform the elementary operations
directly. Then, it makes the function \L{improve_pivot_rec} tail-recursive, which
can have a good impact on performance.

The flip side is that it is slightly more difficult to express and manipulate
formally the link between the matrices taken as arguments and those returned by
the function. Indeed, the specification of this function involves inverses of
transition matrices:

\begin{lstlisting}[language=SSR]
Inductive improve_pivot_rec_spec m n P M Q :
  'M_(1 + m) * 'M_(1 + m,1 + n) * 'M_(1 + n) -> Type :=
  ImprovePivotRecSpec P' M' Q' of
       P^-1 *m M *m Q^-1 = P'^-1 *m M' *m Q'^-1
    & (forall i j, M' 0 0 %| M' i j)
    & (forall i, M' i 0 = M' 0 0)
    & M' 0 0 %| M 0 0
    & P' \in unitmx
    & Q' \in unitmx : improve_pivot_rec_spec P M Q (P',M',Q').
\end{lstlisting}
% \begin{lstlisting}[language=SSR]
% Inductive improve_pivot_rec_spec m n L M R :
%   'M_(1 + m) * 'M_(1 + m,1 + n) * 'M_(1 + n) -> Type :=
%   ImprovePivotRecSpec L' A R' of
%     L^-1 *m M *m R^-1 = L'^-1 *m A *m R'^-1
%     & (forall i j, A 0 0 %| A i j)
%     & (forall i, A i 0 = A 0 0)
%     & A 0 0 %| M 0 0
%     & L' \in unitmx & R' \in unitmx
%   : improve_pivot_rec_spec L M R (L',A,R').
% \end{lstlisting}
%
The statement above can be read as follows: given three matrices \L{P},
\L{M} and \L{Q}, a triple \L{(P,M',Q')} satisfies the specification if applying
to \L{M} the inverse of elementary operations represented by the initial
transition matrices \L{P} and \L{Q} gives the same result as applying the
inverses of the transition matrices \L{P'} and \L{Q'} to \L{M'}.

The correctness lemma of the function \L{improve_pivot_rec} states that for an
initial matrix \L{M} whose upper-left coefficient is nonzero and has a norm
smaller than a natural number \L{k}, and for invertible matrices \L{P} and
\L{Q}, the triple returned by \L{improve_pivot_rec k P M Q} satisfies the
specification represented by the inductive type \L{improve_pivot_rec_spec}:

\begin{lstlisting}[language=SSR]
Lemma improve_pivot_recP k m n (P : 'M_(1 + m)) (M : 'M_(1 + m,1 + n)) Q :
  enorm (M 0 0) <= k -> M 0 0 != 0 ->
  P \in unitmx -> Q \in unitmx ->
  improve_pivot_rec_spec P M Q (improve_pivot_rec k P M Q).
\end{lstlisting}
% \begin{lstlisting}[language=SSR]
% Lemma improve_pivot_recP k m n (L : 'M_(1 + m)) (M : 'M_(1 + m, 1 + n)) R :
%   enorm (M 0 0) <= k -> M 0 0 != 0 ->
%   L \in unitmx -> R \in unitmx ->
%   improve_pivot_rec_spec L M R (improve_pivot_rec k L M R).
% \end{lstlisting}
%
Initially, we call the function \L{improve_pivot_rec} with identity transition
matrices:
\begin{lstlisting}[language=SSR]
Definition improve_pivot k m n (M : 'M_(1 + m, 1 + n)) :=
  improve_pivot_rec k 1%:M M 1%:M.
\end{lstlisting}
By successive subtractions of the first row from all the others and then by
linear combinations of columns, we get a matrix~$C$:

\[ C =\left[\begin{tabular}{c|ccc}
    $g$ & $0$ & $\cdots$ & $0$ \\
    \hline
    $0$ & & & \\
    $\vdots$ & & $C'$ & \\
    $0$ & & & \\
  \end{tabular}\right]
\]
where $g$ divides all coefficients of $C'$.

The global algorithm computing the Smith normal form proceeds as follows: it
stores the pivot $g$ obtained after the previous step, then divides all
coeficients of $C'$ by $g$ and is applied recursively to the resulting matrix.
Let us pose $k=\min(m,n)$. From the pivots $g_1,\ldots,g_k$ obtained, the final
output of the algorithm is given by the following sequence $d_1,\ldots,d_k$:
\[
  d_1,d_2,\ldots,d_k = g_1,g_1 g_2,\ldots,\prod_{i=1}^k g_i
\]
The Smith normal form of the original matrix is then the following diagonal
matrix of size $m \times n$:
\[
  \begin{bmatrix}
    d_1 \\
    & d_2 \\
    & & \ddots \\
    & & & d_k \\
    & & & & 0 \\
    & & & & & \ddots \\
    & & & & & & 0
  \end{bmatrix}
\]
%
% TODO: update to make coherent w.r.t. improve_pivot def
This global procedure is implemented by the function \L{Smith}~:
\begin{lstlisting}[language=SSR,numbers=left,firstnumber=1]
Fixpoint Smith {m n} : 'M[R]_(m,n) -> 'M[R]_(m) * seq R * 'M[R]_(n) :=
  match m, n return 'M[R]_(m, n) -> 'M[R]_(m) * seq R * 'M[R]_(n) with
  | _.+1, _.+1 => fun M : 'M[R]_(1 + _, 1 + _) =>
      if find_pivot M is Some (i, j) then
      let a := M i j in let M := xrow i 0 (xcol j 0 M) in
      let: (P,M,Q) := improve_pivot (enorm a) M in
      let a  := M 0 0 in
      let u  := dlsubmx M in let v := ursubmx M in
      let v' := map_mx (fun x => odflt 0 (x %/? a)) v in
      let M  := drsubmx M - const_mx 1 *m v in
      let: (P', d, Q') := Smith (map_mx (fun x => odflt 0 (x %/? a)) M) in
      (lift0_mx P' *m block_mx 1 0 (- const_mx 1) 1 *m (xcol i 0 P),
       a :: [seq x * a | x <- d],
      (xrow j 0 Q) *m block_mx 1 (- v') 0 1 *m lift0_mx Q')
    else (1%:M, [::], 1%:M)
  | _, _ => fun M => (1%:M, [::], 1%:M)
  end.
\end{lstlisting}
% \begin{lstlisting}[language=SSR,numbers=left,firstnumber=1]
% Fixpoint Smith {m n} : 'M_(m,n) -> 'M_(m) * seq R * 'M_(n) :=
%   match m, n return 'M_(m, n) -> 'M_(m) * seq R * 'M_(n) with
%   | _.+1, _.+1 => fun A : 'M_(1 + _, 1 + _) =>
%     if find_pivot A is Some (i, j) then
%       let a := A i j in let A := xrow i 0 (xcol j 0 A) in
%       let: (L,A,R) := improve_pivot (enorm a) A in
%       let a := A 0 0 in
%       let u := dlsubmx A in let v := ursubmx A in
%       let v' := map_mx (fun x => odflt 0 (x %/? a)) v in
%       let A := drsubmx A - const_mx 1 *m v in
%       let: (L', d, R') := Smith (map_mx (fun x => odflt 0 (x %/? a)) A) in
%       (lift0_mx L' *m block_mx 1 0 (-const_mx 1) 1%:M *m (xcol i 0 L),
%       a :: [seq x * a | x <- d],
%       (xrow j 0 R) *m block_mx 1 (-v') 0 1%:M *m lift0_mx R')
%     else (1%:M, [::], 1%:M)
%   | _, _ => fun A => (1%:M, [::], 1%:M)
%   end.
% \end{lstlisting}
If \L{M} has type \L{'M[R]_n} then \L{lift0_mx M = block_mx 1 0 0 M}
of type \L{'M[R]_(1 + n)}. The notation \L{[seq f x | x <- xs]} is
like a list comprehension in \Haskell{} and means \L{map f xs}.

The function \L{Smith} takes as argument a matrix and returns a
sequence made of the nonzero diagonal coefficients of its Smith form,
as well as the corresponding transition matrices. The first step
(lines~4 and~5) consists in searching for a nonzero pivot in the whole
matrix and moving it in the upper-left position. If no
pivot is found, all the coefficients are zero and an empty sequence
is therefore returned. Otherwise, the function \L{improve_pivot}
defined previously is called (line~6), then some elementary row
operations are performed (lines~8 to~10) to get a matrix of the shape
of the matrix $C$ shown above. The bottom-right submatrix is then
divided by the pivot and a recursive call is performed (line~11). The
sequence of coefficients and transition matrices obtained are then
updated (lines~12 to~14).

We have stated and proved the following correctness lemma:
\begin{lstlisting}[language=SSR]
Lemma SmithP (m n : nat) (M : 'M_(m,n)) : smith_spec M (Smith M).
\end{lstlisting}
Using this we have instantiated the structure of elementary divisor
rings on Euclidean domains.

\subsection{Extension to principal ideal domains}\label{sec:smith-pid}

We mentioned in \sec{sec:dvdrings} that constructive principal ideal domains
were Bézout domains with a well-founded divisibility relation. Well-foundedness
is defined in \Coq{}'s standard library using an accessibility
predicate~\cite{nordstrom:terminating}:

\begin{lstlisting}
Inductive Acc (A : Type) (R : A -> A -> Prop) (x : A) : Prop :=
  Acc_intro : (forall y : A, R y x -> Acc R y) -> Acc R x.
\end{lstlisting}

The idea is that all objects of the inductive type \L{Acc} have to be
built by a finite number of applications of the constructor
\L{Acc_intro}. Hence, for any \L{a} such that \L{Acc R a}, all chains
$(\mathtt{x}_n)$ such that $\mathtt{R}\ \mathtt{x}_{n+1}\ \mathtt{x}_n$ and
$\mathtt{x}_0 = \mathtt{a}$ have to be finite.
 Note however that there can be infinitely many elements \L{x} such that
\L{R x a}. Using this definition of accessibility, we can now state that a
relation over a type \L{A} is well-founded if all elements in \L{A} are
accessible:

\begin{lstlisting}
Definition well_founded (A : Type) (R : A -> A -> Prop) :=
  forall a, Acc R a.
\end{lstlisting}

Remember that in the previous section, we used the hypothesis that the ring of
coefficients was Euclidean when we computed an a priori bound on the number of
steps the function \L{improve_pivot} needed to perform. To extend the algorithm
to principal ideal domains, we replace the recursion on this bound with a
well-founded induction on the divisibility relation.

\begin{lstlisting}
Fixpoint improve_pivot_rec m n (P : 'M_(1 + m)) (M : 'M_(1 + m, 1 + n))
         (Q : 'M_(1 + n)) (k : Acc (@sdvdr R) (M 0 0)) :
         'M_(1 + m) * 'M_(1 + m, 1 + n) * 'M_(1 + n) :=
    match k with Acc_intro IHa =>
      if find1P M (M 0 0) is Pick i Hi then
        let Ai0 := M (lift 0 i) 0 in
        let P := Bezout_step (M 0 0) Ai0 P i in
        improve_pivot_rec P Q (IHa _ (sdvd_Bezout_step Hi))
      else
        let u  := dlsubmx M in let vM := ursubmx M in let vP := usubmx P in
        let u' := map_mx (fun x => 1 - odflt 0 (x %/? M 0 0)) u in
        let P  := col_mx (usubmx P) (u' *m vP + dsubmx P) in
        let A  := block_mx (M 0 0)%:M vM
                           (const_mx (M 0 0)) (u' *m vM + drsubmx M) in
        if find2P A (M 0 0) is Pick (i,j) Hij then
          let A := xrow 0 i A in
          let P := xrow 0 i P in
          let a := A 0 0 in
          let A0j := A 0 (lift 0 j) in
          let Q := (Bezout_step a A0j Q^T j)^T in
          improve_pivot_rec P Q (IHa _ (sdvd_Bezout_step2 Hij))
        else (P, A, Q)
    end.
\end{lstlisting}
The main difference with the function \L{improve_pivot} defined in
\sec{sec:euclidean} is that we need to prove that the upper-left element of the
matrix on which we make the recursive call is strictly smaller than the one of
the original matrix. To build these proofs, we use the functions \L{find1P} and
\L{find2P} which have more expressive (dependent) types than their counterparts
\L{find1} and \L{find2} that we used previously. They return not only an element
of the matrix given as argument, but also a proof that the pivot does not divide
this element.

This proof is then used to show that the upper-left coefficient of the matrix
decreases, thanks to the following two lemmas:

\begin{lstlisting}
Lemma sdvd_Bezout_step m n (M : 'M_(1 + m,1 + n)) (k : 'I_m) :
  ~~ (M 0 0 %| M (lift 0 k) 0) ->
  (Bezout_step (M 0 0) (M (lift 0 k) 0) M k) 0 0 %<| M 0 0.

Lemma sdvd_Bezout_step2 m n i j u' vM (M : 'M[R]_(1 + m, 1 + n)) :
  let B : 'M_(1 + m, 1 + n) :=
    block_mx (M 0 0)%:M vM (const_mx (M 0 0)) (u' *m vM + drsubmx M) in
  let C := xrow 0 i B in
  ~~ (M 0 0 %| B i (lift 0 j)) ->
  (Bezout_step (C 0 0) (C 0 (lift 0 j)) C^T j)^T 0 0 %<| M 0 0.
\end{lstlisting}
Now, to define the \L{improve_pivot} function, we use the hypothesis
\L{sdvdr_wf} that the divisibility relation is well-founded:

\begin{lstlisting}
Definition improve_pivot m n (M : 'M_(1 + m, 1 + n)) :=
  improve_pivot_rec 1 1 (sdvdr_wf (M 0 0)).
\end{lstlisting}
The function \L{Smith} of \sec{sec:euclidean} is essentially unchanged, the only
difference being that we removed the first argument of \L{improve_pivot} (which
was an a priori bound on the number of steps of \L{improve_pivot_rec}).

We have shown how to compute the Smith normal form on Euclidean domains and more
generally on principal ideal domains. In the next section, we will explain how
to develop a constructive theory of linear algebra based on the existence of
such an algorithm.

%% Local Variables:
%% ispell-local-dictionary: "english"
%% mode: latex flyspell
%% TeX-master: "main"
%% End:

\section{Elementary divisor rings}\label{sec:edr}
The goal of this section is to develop some theory about linear
algebra over elementary divisor rings and discuss the formalization of
the classification theorem for finitely presented modules over these
rings.%  The classification theorem involves proving that the Smith
% normal form is unique up to multiplication by units if the underlying
% ring has a $\gcd$ operation.

\subsection{Linear algebra over elementary divisor rings}\label{sec:edrlinalg}
One of the key operations in linear algebra is to compute solutions to
systems of equations. A suitable algebraic setting for doing so is
rings where every finitely generated ideal is finitely presented.
These rings are called coherent:

\begin{defi}
  A ring is \textbf{coherent} if for any matrix $M$ it is possible to
  compute a matrix $L$ such that:
    \[
    XM = 0\  \leftrightarrow \ \exists Y. \, X = YL
    \]
\end{defi}\medskip

\noindent This means that $L$ generates the module of solutions of $XM=0$, \ie{}
that $L$ generates the kernel of $M$. The notion of coherent rings is
usually not mentioned in classical presentations of algebra since
Noetherian rings are automatically coherent, but in a computationally
meaningless way. It is however a fundamental notion, both
conceptually~\cite{LomQui,Mines} and
computationally~\cite{barakat,homalg}. Coherent rings have previously
been represented in \Coq~\cite{Coquand:2012:CSD:2428345.2428366} so we
will not discuss the details of the formalization here. Instead we
show that elementary divisor rings are coherent.

Let $M$ be a $m \times n$ matrix with coefficients in an elementary
divisor ring. There are invertible matrices $P$ and $Q$ such that
$PMQ = D$ where $D$ is a diagonal matrix in Smith normal form. The
rank of $M$, denoted $r(M)$, is the number of nonzero elements of
$D$. The kernel of $M$ can be computed by:
\[
\ker(M) = (I_m - I_{r(M)})P
\]
where $I_m$ is a $m \times m$ identity matrix and $I_{r(m)}$ is a $m
\times m$ partial identity matrix with $r(M)$ ones on the diagonal and
then zeros. The idea behind this definition is that:

\[
\begin{bmatrix}
  0      &        &         &        &        &        \\
         & \ddots &         &        & 0      &        \\
         &        & 0       &        &        &        \\
         &        &         & 1      &        &        \\
         & 0      &         &        & \ddots &        \\
         &        &         &        &        & 1      \\
\end{bmatrix}
\begin{bmatrix}
  d_1    &        &         &        &        &        \\
         & \ddots &         &        & 0      &        \\
         &        & d_k     &        &        &        \\
         &        &         & 0      &        &        \\
         & 0      &         &        & \ddots &        \\
         &        &         &        &        & 0      \\
\end{bmatrix} = 0
\]
So $\ker(M)MQ = 0$ and since $Q$ is invertible, we have $\ker(M)M=0$. We can
implement the rank operator and state its correctness by:
\begin{lstlisting}
Definition mxrank m n (M : 'M[R]_(m,n)) :=
  let: (P,d,Q) := smith M in size [seq x <- d | x != 0 ].

Definition kermx m n (M : 'M[R]_(m,n)) : 'M[R]_m :=
  let: (P,d,Q) := smith M in copid_mx (mxrank M) *m P.

Lemma kermxP m n (M : 'M[R]_(m,n)) (X : 'rV[R]_m) :
  reflect (exists Y : 'rV[R]_m, X = Y *m kermx M) (X *m M == 0).
\end{lstlisting}
where \L{copid_mx} corresponds to the partial identity matrix. The
\L{reflect} statement should be read as: the boolean equality %
\L{X *m M == 0} holds if and only if there exists \L{Y : 'rV[R]_m}
such that \L{X = Y *m kermx M}.

An algorithm computing the cokernel of a matrix can be implemented in
a similar fashion. This way we have implemented a small library
inspired by the one on matrix algebra for fields of
\ssr~\cite{gonthier_point-free_2011}, but based on Smith normal form
instead of Gaussian elimination.

Another important notion in constructive algebra is strongly discrete
rings:

\begin{defi}
  A ring is \textbf{strongly discrete} if membership in finitely generated
  ideals is decidable and if whenever $x \in (x_1,\dots,x_n)$, there exists
  $y_1,\dots,y_n$ such that $x = \sum_i x_i y_i$.
\end{defi}

If a ring is both coherent and strongly discrete it is not only
possible to solve homogeneous systems of equations but also arbitrary
systems of the kind $XM = B$ where $X$ is a $m \times n$ matrix, $M$ a
$n \times k$ matrix and $B$ a nonzero $m \times k$ matrix.

It is easy to see that Bézout domains are strongly discrete as any
finitely generated ideal is principal. To test if $x \in
(a_1,\dots,a_n)$ first compute a principal ideal $(g)$ equivalent to
$(a_1,\dots,a_n)$ and then test if $g \mid x$. If this is the case we
may construct the witness and otherwise we know that $x \notin
(a_1,\dots,a_n)$.

It is also straightforward to prove that any elementary divisor ring
is a Bézout domain. Given $a,b \in R$ we can compute the Smith normal
form of a row matrix containing $a$ and $b$. This gives us an
invertible $1 \times 1$ matrix $P$, an invertible $2 \times 2$ matrix
$Q$, and $g \in R$ such that:
\[
P\begin{bmatrix} a & b \\ \end{bmatrix} Q = \begin{bmatrix} g & 0 \\ \end{bmatrix}
\]
As $P$ and $Q$ are invertible we get that $g$ is the greatest common
divisor of $a$ and $b$. The Bézout coefficients are then found by
performing the matrix multiplications on the left-hand side of the
equality. Hence we get that elementary divisor rings are not only
coherent but also strongly discrete.

In \sec{sec:kaplansky} we consider extensions to Bézout domains that
make them elementary divisor rings and hence form a good setting for
doing linear algebra. The next subsection shows that the existence of
an algorithm for computing the Smith normal form makes finitely
presented modules over elementary divisor rings especially
well-behaved.

\subsection{Finitely presented modules over elementary divisor rings}
Recall that a module is said to be finitely presented if it can be
described using a finite set of generators and a finite set of
relations among these. A convenient way to express this is:

\begin{defi}
  An $R$-module $\mathcal{M}$ is \textbf{finitely presented} if there
  is an exact sequence:
  \begin{center}
    \begin{tikzcd}
      R^{m_1} \arrow{r}{M} & R^{m_0} \arrow{r}{\pi} & \mathcal{M} \arrow{r} & 0
    \end{tikzcd}
  \end{center}
\end{defi}\smallskip

\noindent This means that $\pi$ is a surjection and $M$ a matrix representing
the $m_1$ relations among the $m_0$ generators of the module
$\mathcal{M}$. Another way to think of $\mathcal{M}$ is as the
cokernel of $M$, that is, $\mathcal{M} \simeq \coker(M) = R^{m_0} /
\im(M)$. So a module has a finite presentation if it can be expressed
as the cokernel of a matrix. As all information of finitely presented
modules is contained in its presentation matrix we get that all
algorithms on finitely presented modules can be described by
manipulating the presentation matrices
\cite{Decker2006,Greuel:2007:SIC:1557288,LomQui}.

A morphism $\varphi$ between finitely presented modules $\mathcal{M}$
and $\mathcal{N}$ given by presentations:
\begin{center}
  \begin{tikzcd}[column sep=small]
    R^{m_1} \arrow{r}{M} & R^{m_0} \arrow{r} & \mathcal{M} \arrow{r}
    & 0 & R^{n_1} \arrow{r}{N} & R^{n_0} \arrow{r} & \mathcal{N}
    \arrow{r} & 0
  \end{tikzcd}
\end{center}
is represented by a $m_0 \times n_0$ matrix $\varphi_G$ and a $m_1
\times n_1$ matrix $\varphi_R$ such that the following diagram
commutes:
\begin{center}
  \begin{tikzcd}
    R^{m_1}     \arrow{r}{M} \arrow{d}{\varphi_R} &
    R^{m_0}     \arrow{r}    \arrow{d}{\varphi_G} &
    \mathcal{M} \arrow{r}    \arrow{d}{\varphi}   & 0 \\
    R^{n_1}     \arrow{r}{N} & R^{n_0} \arrow{r}  & \mathcal{N} \arrow{r} & 0
  \end{tikzcd}
\end{center}
The intuition why two matrices are needed is that the morphism affects
both the generators and relations of the modules, hence the
names~$\varphi_G$ and~$\varphi_R$. In this paper we adopt the \ssr{}
convention that composition is read in diagrammatic order (\ie{} from
left to right) when writing equations obtained from commutative
diagrams. This means that the equation related to the above diagram is
written $M \varphi_G = \varphi_R N$.

In order for us to be able to compute kernels of morphisms we need to
assume that the underlying ring is coherent so that we can solve
systems of equations involving the underlying matrices. If the
underlying ring is also strongly discrete, it is possible to
represent morphisms using only~$\varphi_G$ and a proof that $\exists
X. XN = M\varphi_G$ as any system of equations of the kind $XM=B$ is
solvable. Two of the authors have previously~\cite{fpmods} formalized
finitely presented modules over coherent and strongly discrete rings
in \Coq{} which provides a basis for this part of the formalization.

It is in general not possible to decide if two finitely presented
modules are isomorphic or not. However, if the underlying ring is an
elementary divisor ring, it becomes possible. Indeed, let $R$ be an
elementary divisor ring and $M$ be a $m_1 \times m_0$ matrix
presenting an $R$-module $\mathcal{M}$. As $M$ is equivalent to a
diagonal matrix $D$, there are invertible matrices $P$ and $Q$ such
that $M Q = P^{-1} D$. This gives a commutative diagram:
\begin{center}
  \begin{tikzcd}[column sep=large, row sep=large]
    R^{m_1}     \arrow{r}{M} \arrow{d}{P^{-1}} & R^{m_0} \arrow{r} \arrow{d}{Q} & \mathcal{M} \arrow{r} \arrow{d}{\varphi}   & 0 \\
    R^{m_1}     \arrow{r}{D}                   & R^{m_0} \arrow{r}              & \mathcal{D} \arrow{r} & 0
  \end{tikzcd}
\end{center}
We can further prove that $\varphi$ is an isomorphism as $P$ and $Q$
are invertible, and hence get that $\mathcal{M} \simeq \mathcal{D}
\simeq \coker(D)$. Now, since $D$ is a diagonal matrix with nonzero
elements $d_1,\dots,d_n \in R$ on the diagonal, we get that:

\begin{equation}\label{decomposition}
\mathcal{M} \simeq R^{m_0 - n} \oplus R/(d_1) \oplus \dots \oplus R/(d_n)
\end{equation}
with the additional property that $d_i \mid d_{i+1}$ for all $1
\leqslant i < n$. Note that if $d_i$ is a unit then $R/(d_i) \simeq
0$. This means that the theory of finitely presented modules over
elementary divisor rings~$R$ is particularly well-behaved as any
finitely presented $R$-module~$\mathcal{M}$ can be decomposed into a
direct sum of a free module and cyclic modules. This is the first part
of the classification theorem for finitely presented modules over
elementary divisor rings, the second part is the fact that the $d_i$
are unique up to multiplication by units which makes the decomposition
unique.

The uniqueness part is also necessary in order to get a decision
procedure for the isomorphism of finitely presented modules over
elementary divisor rings. So far we only know that any module may be
decomposed as above, but there is, a priori, no reason why two
isomorphic modules should have related decompositions.

In the next section we will see that the Smith normal form is unique
up to multiplication by units if the underlying ring has a $\gcd$
operation, which in turn completes the classification theorem and
gives us a decision procedure for module isomorphism.

\subsection{Uniqueness of the Smith normal form}\label{subsec:uniqsnf}

% \gc{It is strange to have this part here. I propose this reorganisation :
%     \begin{itemize}
%       \item First, present the Smith algorithm for pid.
%       \item Second, explain that Euclidean rings are pid but in this case
%         we have an more efficient algorithm and explain it.
%       \item Third, explain unicity of the Smith normal form.
%     \end{itemize} }

The formal proof that the Smith normal form is unique up to
multiplication by units presented here is based
on~\cite{cano:hal-00779376}. In order to formalize this proof we need
to represent minors~(determinants of submatrices) in \Coq{}. This
notion was defined in a previous work on formalizing the Sasaki-Murao
algorithm computing the characteristic polynomial of a
matrix~\cite{JFR2615}. With the~\ssr{} definition of matrices it is
easy to give a definition of submatrices (denoted by $M(f,g)$) and
minors:

\begin{lstlisting}
Definition submatrix m n p q (f : 'I_p -> 'I_m) (g : 'I_q -> 'I_n)
  (M : 'M[R]_(m,n)) : 'M[R]_(p,q) :=
  \matrix_(i < p, j < q) M (f i) (g j).

Definition minor m n p (f : 'I_p -> 'I_m) (g : 'I_p -> 'I_n)
  (M : 'M[R]_(m,n)) : R := \det (submatrix f g M).
\end{lstlisting}

For example, the rows (resp. columns) of the matrix $M(f,g)$ are the
rows (resp. columns) $f(0),f(1),...$ (resp. $g(0),g(1),...$) of
$M$. It would be natural to define submatrices only when
\L{f} and \L{g} are strictly increasing, however this is not necessary
as many theorems are true for arbitrary functions. We denote \L{p} in
the definition of \L{minor} above as the order of the minor, that is,
a minor of order $p$ is the determinant of a submatrix of dimension $p
\times p$.

The key result in order to prove the uniqueness theorem for the Smith
normal form is that the product of the $k$ first elements of the
diagonal in the Smith normal form is associated to the $\gcd$ of the
minors of order $k$ of the original matrix. More precisely, let $M$ be
the original matrix and $d_i$ the $i$:th element of the diagonal in
the Smith normal form of~$M$, also let $\vec{m}_k$ be the minors of
order~$k$ of~$M$, then the statement is:

\[
  \prod_{i=1}^k d_i = \gcd(\vec{m}_k)
\]
Using the big operators library of \ssr~\cite{BGBP} this can be
expressed compactly as:

\begin{lstlisting}
Lemma Smith_gcdr_spec :
  \prod_(i < k) d`_i %= \big[gcdr/0]_f \big[gcdr/0]_g minor f g M.
\end{lstlisting}

The order of the minors that we consider are given by the types of
\L{f} and \L{g}. For the sake of readability, we have omitted these
types.

The first step in proving this is by showing that it holds for the Smith normal
form of~$M$, namely the diagonal matrix~$D$. Since it is a diagonal matrix, the
only nonzero minors of order~$k$ are the determinants of diagonal matrices of
dimension $k \times k$, that are products of~$k$ elements of the diagonal of
$D$. Also, since each element of the diagonal divides the next one, the greatest
common divisor of the minors of order $k$ is the product of the $k$ first
elements of the diagonal. For example, if the diagonal is $(a,b,c)$ with $a \mid
b$ and $b \mid c$ then $\gcd(ab,bc,ac)= ab$.

The next step is to prove that the $\gcd$ of the minors of order $k$
of~$M$ are associated to the $\gcd$ of the minors of $D$ (which we
already know is associated to the product of the elements on the
diagonal). To prove this it suffices to show that these two divide
each other, as the proofs in both directions are very similar we only
show that the $\gcd$ of the minors of order $k$ of $M$ divides the
$\gcd$ of the minors of order $k$ of $D$.

By definition, $x$ divides $\gcd(\vec{y})$ if and only if $x$
divides every $y$ in $\vec{y}$. So we must show that the $\gcd$ of
the minors of order $k$ of $M$ divides each minor of order $k$ of the
diagonal matrix $D$. Now, there are invertible matrices $P$ and $Q$
such that $PMQ = D$. Hence we must show that $\gcd(\vec{m}_k)$ divides
$\det((PMQ)(f,g))$ for all $f$ and $g$. The right-hand side is the
determinant of a product of matrices of different sizes whose product
is square, which can be simplified with the Binet-Cauchy formula:
\[
\det(MN) = \sum_{I \in \mathcal{P} (\{1,\ldots, l\}) \atop \#|I| = k}
  \det(M_I)\det(N_I)
\]
where $M$ is a $k \times l$ matrix and $N$ is a $l \times k$
matrix. $M_I$ (resp.~$N_I$) is the matrix of the $k$ columns
(resp.~rows) with indices in $I$. 

The formalization of this formula
builds on the work in \cite{JFR2615} and follows Zeng's proof
presented in \cite{Zeng199379}. Note that the standard determinant
identity for products of square matrices of the same size follows as a
special case of the above formula. Once again the theorem can be
expressed compactly using the big operators of~\ssr{}:

\begin{lstlisting}
Lemma BinetCauchy :
  \det (M *m N) = \sum_(f : {ffun 'I_k -> 'I_l} | strictf f)
                          ((minor id f M) * (minor f id N)).
\end{lstlisting}

Here the sum is taken over all strictly increasing functions from
$\{1,\ldots,k\}$ to $\{1,\ldots,l\}$. We require the functions to be
strictly increasing so that the minors that we consider in the sum correspond
to the mathematical concept of minor.
%% The problem here is not that some terms of the sum are null (since zero is a
%% neutral element of addition). The problem is that if f corresponds to a
%% permutation then this add a term not necessary null in the sum and this
%% falsify the result.
%%we do not pick out the same column or row
%%twice making some of the minors zero.

This theorem makes it possible for us to transform $\det((PMQ)(f,g))$
to a sum of minors and, once again, it suffices to show that
$\gcd(\vec{m}_k)$ divides each of the summands. Hence, after some
simplifications, we must show that for all \L{h} and \L{i} we have:

\begin{lstlisting}
  \big[gcdr/0]_f \big[gcdr/0]_g minor f g M %| minor h i M
\end{lstlisting}
which is true by definition of the $\gcd$. Note that it is not
necessary to require that \L{f} and \L{g} are strictly
increasing. Indeed, if they are not, there are two cases:

\begin{itemize}
\item Either \L{f} or \L{g} is not injective and so \L{minor f g M = 0}.

\item If both \L{f} and \L{g} are injective there exist permutations
  \L{r} and \L{s} such that \L{f' = f \o r} and \L{g' = g \o s} are
  strictly increasing. As the permutation of rows or columns of a
  matrix just leads to the determinant being multiplied by the
  signature of the permutation we get \L{minor f g M %= minor f' g' M}.
\end{itemize}
But for all $a$ we have $\gcd(a,0) = a$ and $\gcd(a,a) = a$, so in
each case the terms corresponding to the minors obtained from not
strictly increasing \L{f} and \L{g} does not change the value of the
$\gcd$ of the minors.

Now if the above result is applied with $k = 1$, the uniqueness of the first
diagonal element is proved, and then by induction all of the diagonal elements
are showed to be unique (up to multiplication by units). This means that for any
matrix $M$ equivalent to a diagonal matrix $D$ in Smith normal form, each of the
diagonal elements of the Smith normal form of $M$ will be associate to the
corresponding diagonal element in $D$. The uniqueness of the Smith normal form
is expressed formally as follows:

\begin{lstlisting}
Lemma Smith_unicity m n (M : 'M[R]_(m,n)) (d : seq R) :
  sorted %| d -> equivalent M (diag_mx_seq m n d) ->
  forall i, i < minn m n -> (smith_seq M)`_i %= d`_i.
\end{lstlisting}
% \begin{lstlisting}
% Lemma Smith_unicity m n (A : 'M[R]_(m,n)) (s : seq R) :
%   sorted %| s -> equivalent A (diag_mx_seq m n s) ->
%   forall i, i < minn m n -> s`_i %= (Smith_seq A)`_i.
% \end{lstlisting}
Hence we have proved that the Smith normal form is unique up to
multiplication by units. This gives a test to know if two matrices are
equivalent. Indeed, since the Smith normal form of a matrix is
equivalent to it, two matrices are equivalent if and only if they have
the same normal form. Moreover, we know that the decomposition in
equation~\eqref{decomposition} is unique up to multiplication by
units. Hence we get an algorithm for deciding if two finitely presented
modules are isomorphic or not: compute the Smith normal form of the
presentation matrices and then test if they are equivalent up to
multiplication by units.

This concludes the classification theorem for finitely presented
modules over elementary divisor rings. It can be seen as a
constructive version of the classification theorem for finitely
generated modules over principal ideal domains. Classical proofs of
this use the fact that a principal ideal domain $R$ is Noetherian
which implies that any $R$-module is coherent,~\ie{} that any finitely
generated module is also finitely presented. But this proof has no
computational content (see exercise 3 in chapter III.2
of~\cite{Mines}), so instead we have to restrict to finitely presented
modules. In \sec{sec:smith-pid} we showed that (constructive)
principal ideal domains are elementary divisor rings which gives us
the classical result in the case of finitely presented modules. In the
next section we will prove that more general classes of rings than
principal ideal domains are elementary divisor rings which gives more
instances of the classification theorem.

%%% Local Variables:
%%% mode: latex
%%% TeX-master: "main"
%%% End:

\section{Extensions to Bézout domains that are elementary divisor
  rings}\label{sec:kaplansky}

As mentioned in the introduction, it is an open problem whether all
Bézout domains are elementary divisor rings or not. In order to
overcome this, we study different properties that we can extend Bézout
domains with to make them elementary divisor rings. The properties we
define and discuss in this section are:

\begin{enumerate}
  \item Adequacy (\ie{} the existence of a $\gdco$ operation);
  \item Krull dimension $\leq 1$;
  \item Strict divisibility is well-founded (constructive principal
    ideal domains).
\end{enumerate}
We have already considered the last one of these in \sec{sec:algo},
but here we formalize an alternative proof that constructive principal
ideal domains are elementary divisor rings, using a reduction due to
Kaplansky~\cite{Kaplansky}. It consists in first simplifying the
problem of computing Smith normal form for $m \times n$ matrices to
the $2 \times 2$ case and then showing that any $2 \times 2$ matrix
has a Smith normal form if and only if the ring satisfies the
``Kaplansky condition''. This means that it suffices for us to prove
that the three different extensions all imply this condition in order
to show that they are elementary divisor rings.

\subsection{The Kaplansky condition}
The reduction of the computation of Smith normal form of arbitrary $m
\times n$ matrices to $2 \times 2$ matrices is done by extracting an
algorithm from the proof of theorem 5.1 in \cite{Kaplansky}. The
formalization is done by first implementing this algorithm, called
\L{smithmxn}, computing the Smith normal form of arbitrary sized
matrices assuming an operation computing it for $2 \times 2$ matrices
and then proving that this algorithm satisfies \L{smith_spec}:

\begin{lstlisting}
Lemma smithmxnP :
  forall (smith2x2 : 'M[R]_2 -> 'M[R]_2 * seq R * 'M[R]_2),
  (forall (M : 'M[R]_2), smith_spec M (smith2x2 M)) ->
  forall m n (M : 'M[R]_(m,n)), smith_spec M (smithmxn smith2x2 M).
\end{lstlisting}
This algorithm has no assumptions on the underlying ring except that
it is an integral domain. It can be generalized to arbitrary
commutative rings but then we also need to be able to put $1 \times 2$
and $2 \times 1$ matrices in Smith normal form.

Now consider a $2 \times 2$ matrix
\[
\begin{bmatrix}
  a & b \\
  c & d \\
\end{bmatrix}
\]
with coefficients in a Bézout domain. We can compute $g = \gcd(a,c)$
and $a_1$ and $c_1$ such that $a = a_1 g$ and $c = c_1 g$. We also
have $u$ and $v$ such that $ua_1 + vc_1 = 1$. Using this we can form:
\begin{align*}
\begin{bmatrix}
  u    & v  \\
  -c_1 & a_1 \\
\end{bmatrix}
\begin{bmatrix}
  a & b \\
  c & d \\
\end{bmatrix}
&=
\begin{bmatrix}
  ua+vc & ub+vd \\
  -c_1a+a_1c & -c_1b+a_1d \\
\end{bmatrix}\\
&=
\begin{bmatrix}
  ua+vc & ub+vd \\
  0     & -c_1b+a_1d \\
\end{bmatrix}
\end{align*}
So it suffices to consider matrices of the following shape:
\[\begin{bmatrix} a & b \\ 0 & c \\ \end{bmatrix}\]
and without loss of generality we can
assume that $\gcd(a,b,c) = 1$. Now, such a matrix has a Smith normal
form if and only if it satisfies the \textbf{Kaplansky condition}:
for all $a,b,c \in R$ with $\gcd(a,b,c) = 1$ there exist $p,q \in R$
with $\gcd(pa,pb+qc) = 1$.

The interesting step for the reduction is the right to left direction
of the ``if and only if'', so let us sketch how it is proved: assume
that $R$ is a Bézout domain that satisfies the Kaplansky condition and
consider an upper triangular matrix with elements $a$, $b$ and $c$
with $\gcd(a,b,c) = 1$. From the Kaplansky condition we get $p$ and
$q$ such that $\gcd(pa,pb+qc) = 1$. This means that we also have $x_1$
and $y_1$ such that $pax_1 + (pb+qc)y_1 = 1$. By reorganizing this we
get $p(ax_1 + by_1) + qcy_1 = 1$, let $x = ax_1 + by_1$ and
$y = cy_1$. We can form the product:

\[
\begin{bmatrix}
  p  & q \\
  -y & x \\
\end{bmatrix}
\begin{bmatrix}
  a & b \\
  0 & c \\
\end{bmatrix}
\begin{bmatrix}
  x_1 & pb+qc \\
  y_1 & -pa \\
\end{bmatrix}
=
\begin{bmatrix}
  1 & 0 \\
  0 & -ac \\
\end{bmatrix}
\]

In order to formalize this proof we assume that we have an operation
taking~$a$,~$b$ and~$c$ computing~$p$ and~$q$ satisfying the Kaplansky
condition:

\begin{lstlisting}
Variable kap : R -> R -> R -> R * R.

Hypothesis kapP : forall (a b c : R),  gcdr a (gcdr b c) %= 1 ->
  let: (p,q) := kap a b c in coprimer (p * a) (p * b + q * c).
\end{lstlisting}

We then define a function \L{kapW : R -> R -> R -> R * R} to extract
the two witnesses $x_1$ and $y_1$ from above, \ie{} $x_1$ and $y_1$
such that $x_1 p a + y_1 (p b + q c) = 1$. To do this we first prove:

\begin{lstlisting}
Lemma coprimerP (a b : R) :
  reflect (exists (xy : R * R), xy.1 * a + xy.2 * b = 1) (coprimer a b).
\end{lstlisting}
and we can then define a function computing $(x_1,y_1)$ by turning the
existential statement in \L{coprimerP} into a $\Sigma$-type (\ie{} a
dependent pair). More precisely, we have defined it by:

\begin{lstlisting}
Definition kapW a b c : R * R :=
  let: (p,q) := kap a b c in
  if coprimerP (p * a) (p * b + q * c) is ReflectT P
     then projT1 (sig_eqW P) else (0,0).
\end{lstlisting}
%% Where \L{sig_eqW} is the following lemma define in the \ssr{} library:
%% \begin{lstlisting}
%% Lemma sig_eqW (T : choiceType) (vT : eqType) (lhs rhs : T -> vT) :
%%   (exists x, lhs x = rhs x) -> {x | lhs x = rhs x}.
%% \end{lstlisting}
Here \L{sig_eqW} is a function from the \ssr{} library that transforms
our existential statement into a $\Sigma$-type, the first component of
the resulting $\Sigma$-type is then extracted using \L{projT1}. This
is possible because \L{R} is taken to be an~\ssr{} ``choice type'',
\ie{} a type with a choice operator.

Once we have defined \L{kapW}, we can easily write the function
computing Smith normal form of $2 \times 2$ matrices, called
\L{kap_smith}, and prove that it satisfies \L{smith_spec}:

\begin{lstlisting}
Definition kap_smith (M : 'M[R]_2) : 'M[R]_2 * seq R * 'M[R]_2 :=
  let: A := Bezout_step (M 0 0) (M 1 0) M 0 in
  let: a00 := A 0 0 in let: a01 := A 0 1 in let: a11 := A 1 1 in
  let: (d,_,_,_,a,b,c) := egcdr3 a00 a01 a11 in
  if d == 0 then (Bezout_mx (M 0 0) (M 1 0) 0,[::],1%:M) else
    let: (p,q) := kap a b c in
    let: (x1,y1) := kapW a b c in
    let: (x,y) := (a * x1 + y1 * b, c * y1) in
     (mx2 p q (- y) x *m Bezout_mx (M 0 0) (M 1 0) 0,
      map (fun x => d * x) [:: 1; - a * c],
      mx2 x1 (p * b + q * c) y1 (- p * a)).

Lemma kap_smithP (M : 'M[R]_2) : smith_spec M (kap_smith M).
\end{lstlisting}
Here \L{mx2} is a notation to define $2 \times 2$ matrices and
\L{egcdr3} computes the Bézout coefficients for $3$ elements.

We have also formalized the other direction, so for a Bézout domain,
satisfying the Kaplansky condition is equivalent to being an
elementary divisor ring. Hence it suffices to prove that the various
extensions to Bézout domains satisfy the Kaplansky condition in
order to get that they are elementary divisor rings.

\subsection{The three extensions to Bézout domains}

In this section we discuss three extensions to Bézout domains that
imply the Kaplansky condition.

\subsubsection{Adequate domains}\label{sec:kapadequate}

In~\cite{Helmer} Helmer introduced the notion of \textbf{adequate
  domains}. These are Bézout domains where for any~$a,b\in R$, with $b
\neq 0$, there exists~$r \in R$ such that:

\begin{enumerate}
  \item $r \mid b$,
  \item $r$ is coprime with $a$, and
  \item for all non unit $d$ such that $d r \mid b$ we have that $d$
    is not coprime with $a$.
\end{enumerate}

We have proved that this notion is equivalent to having a ``$\gdco$''
function. This function has previously been introduced by one of the
authors in~\cite{ACF} in order to implement quantifier elimination for
algebraically closed fields. It has also other applications in algebra,
see~\cite{Luneburg:1985:LBU:646022.676247}. It takes two elements $a,b
\in R$, with $b \neq 0$, and computes $r$ such that:

\begin{enumerate}
  \item $r \mid b$,
  \item $r$ is coprime with $a$ , and
  \item for all divisors $d$ of $b$ that is coprime to $a$ we have $d
    \mid r$.
\end{enumerate}
This means that $r$ is the greatest divisor of $b$ that is coprime to
$a$. These notions are expressed in \Coq{} as:

\begin{lstlisting}
Inductive adequate_spec (a b : R) : R -> Type :=
  | AdequateSpec0 of b = 0 : adequate_spec a b 0
  | AdequateSpec r of b != 0
                      & r %| b
                      & coprimer r a
                      & (forall d, d * r %| b -> d \isn't a GRing.unit ->
                          ~~ coprimer d a)
                      : adequate_spec a b r.

Inductive gdco_spec (a b : R) : R -> Type :=
  | GdcoSpec0 of b = 0 : gdco_spec a b 0
  | GdcoSpec r of b != 0
                  & r %| b
                  & coprimer r a
                  & (forall d, d %| b -> coprimer d a -> d %| r)
                  : gdco_spec a b r.

Lemma adequate_gdco a b r : adequate_spec a b r -> gdco_spec a b r.
Lemma gdco_adequate a b r : gdco_spec a b r -> adequate_spec a b r.
\end{lstlisting}

We have implemented an algorithm called \L{gdco_kap} that computes $p$
and $q$ in the Kaplansky condition using the $\gdco$ operation. Using
this we have proved:

\begin{lstlisting}
Lemma gdco_kapP (a b c : R) : gcdr a (gcdr b c) %= 1 ->
  let: (p, q) := gdco_kap a b c in coprimer (p * a) (p * b + q * c).
\end{lstlisting}

Using this we can define a function that computes the Smith normal
form for any matrix over an adequate domain:

\begin{lstlisting}
Definition gdco_smith := smithmxn (kap_smith gdco_kap).

Lemma gdco_smithP m n (M : 'M[R]_(m,n)) : smith_spec M (gdco_smith M).
\end{lstlisting}
Hence we get that adequate domains are elementary divisor rings.

\subsubsection{Krull dimension~$\leq 1$}\label{sec:kapkrull1}
The next class of rings we study are Bézout domains of Krull
dimension~$\leq 1$. Classically Krull dimension is defined as the
supremum of the length of all chains of prime ideals, this means that
a ring has Krull dimension $n \in \N$ if there is a chain of
prime ideals:
\[
\mathfrak{p}_0 \subsetneq \mathfrak{p}_1 \subsetneq \dots \subsetneq \mathfrak{p}_n
\]
but no such chain of length $n+1$. For example, a field has Krull
dimension $0$ and any principal ideal domain (that is not a field) has
Krull dimension $1$. This can be defined constructively using an
inductive definition as in~\cite{LomQui}. Concretely an integral
domain $R$ is of Krull dimension~$\leq 1$ if for any $a,u \in R$ there
exists $v \in R$ and $n \in \N$ such that
\[
a \mid u^n (1 - uv)
\]\smallskip

\noindent In order to prove that Bézout domains of Krull dimension~$\leq 1$
are adequate we first prove:

\begin{lstlisting}
Hypothesis krull1 : forall a u, exists m v, a %| u ^+ m * (1 - u * v).

Lemma krull1_factor a b : exists n b1 b2,
  [&& 0 < n, b == b1 * b2, coprimer b1 a & b2 %| a ^+ n].
\end{lstlisting}
This means that given $a$ and $b$ we can compute $n \in \N$
and $b_1,b_2 \in R$ such that $n \neq 0$, $b = b_1 b_2$, $b_1$ is
coprime with $a$ and $b_2 \mid a^n$. If we set $r$ to $b_1$ in the
definition of adequate domains we have to prove:

\begin{enumerate}
  \item $b_1 \mid b_1 b_2$,
  \item $b_1$ is coprime with $a$, and
  \item for all non unit $d$ such that $d b_1 \mid b_1 b_2$ we have
    that $d$ is not coprime with $a$.
\end{enumerate}
The first two are obvious. For the third point, we have to
prove that any non-unit $d$ that divides $b_2$ is not coprime with
$a$. So it suffices to prove that any $d$ coprime with $a$ that
divides $b_2$ is a unit. Now as $n \neq 0$ we get that $d$ is coprime
with $a^n$, but $d \mid b_2$ and $b_2 \mid a^n$ so $d$ must be a
unit. We have formalized this argument in:

\begin{lstlisting}
Lemma krull1_adequate a b : { r : R & adequate_spec a b r }.
\end{lstlisting}
This means that Bézout domains of Krull dimension~$\leq 1$ are adequate and
hence satisfy the Kaplansky condition, which in turn means that they are
elementary divisor rings:

\begin{lstlisting}
Definition krull1_gdco a b := projT1 (krull1_adequate a b).

Definition krull1_smith := gdco_smith krull1_gdco.

Lemma krull1_smithP m n (M : 'M[R]_(m,n)) : smith_spec M (krull1_smith M).
\end{lstlisting}

\subsubsection{Constructive principal ideal domains}\label{sec:kappid}
Finally, we have showed that constructive principal ideal domains are
adequate domains by proving that given $a$ and $b$ we can compute $r$
satisfying \L{gdco_spec}:

\begin{lstlisting}
Lemma pid_gdco (R : pidType) (a b : R) : {r : R & gdco_spec a b r}.
\end{lstlisting}

The construction of the greatest divisor of $a$ coprime to $b$ in a
constructive principal ideal domain is done as in the particular case of
polynomials in~\cite{ACF}. If $\gcd(a,b)$ is a unit, then~$a$ is
trivially the result, otherwise we get~$a'$ by dividing~$a$ by
$\gcd(a,b)$ and we repeat the process with~$a'$ and~$b$. This process
terminates because when $\gcd(a,b)$ is not a unit, $a'$~strictly
divides~$a$ and by our definition of constructive principal ideal domains,
there cannot be an infinite decreasing sequence for strict divisibility.

This way we get an alternative proof that constructive principal ideal
domains are elementary divisor rings:

\begin{lstlisting}
Definition pid_smith := gdco_smith (fun a b => projT1 (pid_gdco a b)).

Lemma pid_smithP m n (M : 'M[R]_(m,n)) : smith_spec M (pid_smith M).
\end{lstlisting}
This proof is simpler in the sense that we first reduce the problem of
computing the Smith normal form to computing the $\gdco$ of two
elements. This way, the part of the proof based on well-founded
recursion is concentrated to \L{pid_gdco} instead of being interleaved
in the algorithm computing the Smith normal form of arbitrary $m
\times n$ matrices.

%% Local Variables:
%% ispell-local-dictionary: "english"
%% mode: latex flyspell
%% TeX-master: "main"
%% End:

\section{Related work}\label{sec:related}
Most proof systems have one or more libraries of formalized linear
algebra. However, the specificity of our work is that it is more general than
the usual study of vector spaces (we do not require scalars to be in a field,
but only in an elementary divisor ring) while still retaining an algorithmic
basis, as opposed to a purely abstract and axiomatized development. In
particular, this work constitutes to our knowledge the first formal verification
of an algorithm for the Smith normal form of matrices.

A fair amount of module theory and linear algebra has been
formalized~\cite{rudnicki:commutative} in \Mizar. But it is based on
classical logic and does not account for underlying algorithmic
aspects. Likewise, a \HOLLight{} library~\cite{harrison:euclidean}
proves significant results in linear algebra and on the topology of
vector spaces, but it is specialized to $\mathbb{R}^n$ and also
classical.

Some other developments focus more on the algebra of vectors and matrices,
without providing support for point-free reasoning on subspaces. Let us
cite~\cite{obua:linear} in \Isabelle, which aims primarily to certify linear
inequalities and \cite{gamboa:arrays,hendrix:matrices} in \ACL, formalizing only
matrix algebra.

In \Coq{} too, older developments focus on the representation of matrices
like~\cite{magaud:matrices}, or classical linear algebra over a field
like~\cite{stein:linear}, based on~\cite{pottier:algebra}. One exception is of
course the more recent work~\cite{gonthier_point-free_2011} we already
mentioned and on which we based this work, extending it from finitely generated
vector spaces to finitely presented modules over elementary divisor
rings.

The authors are also developing a library of computational algebra
called \CoqEAL{} -- the \Coq{} Effective Algebra
Library~\cite{refinements_for_free,coqeal}. It contains many examples
of algorithms from linear algebra like the rank of matrices over
fields and Strassen's matrix multiplication~\cite{coqeal}, the
Sasaki-Murao algorithm for computing the characteristic polynomial of
a matrix over a commutative ring~\cite{JFR2615}, and the kernel of a
matrix over a field~\cite{pershomology}.

Two of the authors have previously formalized the theory of finitely
presented modules in \Coq~\cite{fpmods}, building on a previous
formalization of coherent and strongly discrete
rings~\cite{Coquand:2012:CSD:2428345.2428366} that provides a basis
for a general treatment of matrix algebra. The present work extends
this to the theory of finitely presented modules over elementary
divisor rings, which gives a means for deciding whether two finitely
presented modules are isomorphic or not as described
in~\sec{subsec:uniqsnf}. It also provides concrete instances solving
the basic algorithmic problems underlying the work on finitely
presented modules as elementary divisor rings provides interesting
examples of coherent strongly discrete rings.

There has also been a lot of work on implementing algorithms for
computing the Smith normal form over various rings in computer algebra
systems like \textsc{Axiom}, \textsc{Maple} and \textsc{Magma}.
% \am{add
% proper references, maybe look what has been done in commercial
% systems like mathematica and matlab as well?}.
There are also lots of literature on very efficient algorithms for
computing the Smith normal form over various coefficient rings, see
for
instance~\cite{Dumas200171,Storjohann1996,Storjohann199825,Storjohann1997155,Villard1995269}.
The motivation behind this kind of work is however different from the
work presented here as the focus is on devising very efficient
algorithms without formal proofs of correctness. Because of these the
focus is on specific coefficient rings and not on full
generality. However, using the \CoqEAL{} approach we expect that it
would be both possible and very interesting to extend this work to
also implement more efficient algorithms from computer algebra.

% \begin{itemize}
%   \item Our own work \cite{fpmods} and \cite{Coquand:2012:CSD:2428345.2428366}
%     \md{Cyril and Anders, can you write something on these two?}
%    See e.g. Harrison, Aransay.
%  \item Linear algebra over rings if we find some (classical, constructive) in
%    the same provers.
%  \item Homology in ACL2 and Coq (at least)
% \end{itemize}

%% Local Variables:
%% ispell-local-dictionary: "english"
%% mode: latex flyspell
%% TeX-master: "main"
%% End:

\section{Conclusions and future work}\label{sec:conclusion}
The relationships between the notions introduced in this paper are
depicted in~\fig{fig:diag}. The numbers on the edges denote the
sections in which the different implications and inclusions are
proved:

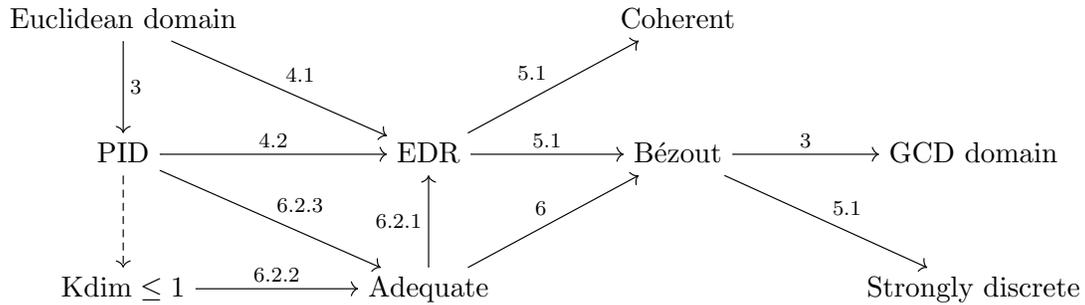
\begin{figure}[h]
  \centering
  \begin{tikzcd}[column sep=large, row sep=large]
     \text{Euclidean domain} \arrow{d}{\ref{sec:dvdrings}} \arrow{dr}{\ref{sec:euclidean}}      & {}                                   & \text{Coherent}                    & {} \\
     \text{PID} \arrow[dashed]{d} \arrow{dr}{\ref{sec:kappid}} \arrow{r}{\ref{sec:smith-pid}} & \text{EDR} \arrow{ur}{\ref{sec:edrlinalg}} \arrow{r}{\ref{sec:edrlinalg}} & \text{Bézout} \arrow{dr}{\ref{sec:edrlinalg}} \arrow{r}{\ref{sec:dvdrings}} & \text{GCD domain} \\
     \text{Kdim} \leq 1 \arrow{r}{\ref{sec:kapkrull1}}                      & \text{Adequate} \arrow{u}{\ref{sec:kapadequate}} \arrow{ur}{\ref{sec:kaplansky}} &                                    & \text{Strongly discrete} \\
  \end{tikzcd}
  \caption{Relationship between the defined notions}
  \label{fig:diag}
\end{figure}
The arrow between PID and Krull dimension $\leq 1$ is dashed because it has not
been formally proved yet. A constructive proof of this can be found
in~\cite{LomQui}. We currently see two options to formalize it: either we try to
develop more extensively the theory of ideals to stick close to the paper
proof, or we expand statements on ideal to statements on elements. Unlike the
former, the latter option would require no further infrastructure, but it is
likely that the size of the proof would explode, as in some proofs where we
already had to talk about elements instead of ideals (\eg{} the lemma
\C{krull1_factor} in the current state of the formalization).

It has been mentioned that $\Z$ and $k[x]$ where $k$ is a field are
the basic examples for all of these rings. Many more examples of
Bézout domains are presented in the chapters on Bézout domains and
elementary divisor rings in~\cite{fuchs2001modules} (for instance,
Bézout domains of arbitrary finite Krull dimension and an example of a
Bézout domain that is not adequate). It would be interesting see which
of these could be done in a constructive setting and formalize them in
order to get more instances than $\Z$ and $k[x]$.

Note that the Kaplansky condition in~\sec{sec:kaplansky} is expressed
using first-order logic. It means that the open problem whether all
Bézout domains are elementary divisor rings can be expressed using
first-order logic. We have formulated the problem this way and applied
various automatic theorem provers in order to try to find a proof that
Bézout domains, alone, and with the two other assumptions (adequacy or
Krull dimension $\leq 1$) are elementary divisor rings. However, none
managed so far.

We have in this paper presented the formalization of many results on
elementary divisor rings. This way we get interesting examples of
coherent strongly discrete rings and concrete algorithms for studying
finitely presented modules. All of the proofs have been performed in a
constructive setting, and except for principal ideal domains, without
chain conditions.

The size of the companion material is approximately 9000~lines of
code.

An important application of this work would be to compute the homology
of chain complexes which provides a means to study properties of
mathematical objects like topological spaces. By computing homology
one associates modules to these kinds of objects, giving a way to
distinguish between them. The Smith normal form of matrices with
coefficients in the ring of integers (denoted by~$\Z$) is at the heart
of the computation of homology as the universal coefficient theorem
for homology~\cite{Hatcher} states that homology with coefficients
in~$\Z$ determines homology with coefficients in any other abelian
group. By developing the theory for more general rings than~$\Z$ it
should be possible to implement and reason about other functors from
homological algebra as well, like for instance cohomology and the Ext
and Tor functors.

%% Local Variables:
%% ispell-local-dictionary: "english"
%% mode: latex flyspell
%% TeX-master: "main"
%% End:

\section*{Acknowledgments}
The authors would like to thank Thierry Coquand and Henri Lombardi for
interesting discussions. The authors are also grateful to Dan Rosén
and Jean-Christophe Filliâtre for helping us explore the Kaplansky
condition using various automatic theorem provers. We would also like
to thank Claire Tête for useful comments on a preliminary version of
the paper. Finally we would like to thank the anonymous reviewers for
their helpful remarks.

\bibliographystyle{abbrvurl}
\bibliography{edr}

\end{document}